\documentstyle[11pt,newpasp,twoside,epsf]{article}
\def\edcomment#1{\iffalse\marginpar{\raggedright\sl#1\/}\else\relax\fi}

\reversemarginpar

\begin{document}

\title{The Physical Conditions and Metal Enrichment of Low-Redshift 
Interstellar and Intergalactic Media: 
The Benefits of High-Resolution Ultraviolet Spectra}

\author{Todd M. Tripp and David V. Bowen}

\affil{Department of Astrophysical Sciences, Princeton University, Princeton, NJ 
08544}

\begin{abstract}
To underscore the value of high spectral resolution for the study of low$-z$ QSO 
absorption lines, we briefly present Space Telescope Imaging Spectrograph 
echelle data that demonstrate how increased resolution leads to dramatic 
improvement in line measurements. We show that even $R = \lambda/\Delta \lambda 
\approx$ 20,000 is insufficient for some measurements. The higher the resolution 
the better, but $R \approx$ 50,000 is adequate for many outstanding questions 
about the IGM and the ISM of galaxies that can be probed using QSO absorption 
lines.
\end{abstract}

\section{Introduction}

QSO absorption lines provide a sensitive probe of
the physical conditions and enrichment of the
diffuse intergalactic medium as well as the
interstellar media of galaxies. It is well know that at low redshifts,
the critical absorption lines are all in the
ultraviolet and must be observed from space. However, 
the need for high spectral resolution is not always
fully appreciated. This poster presented several
examples, from real data obtained with the
Space Telescope Imaging Spectrograph (STIS), of the
benefits of {\it high-resolution} spectroscopy for the
study of QSO absorbers. Some of these examples are reproduced in 
this short contribution to illustrate the importance of high 
spectral resolution as we consider options for more powerful 
telescope/instrument combinations to follow in the footsteps of 
the {\it Hubble Space Telescope}. The data used for these demonstrations were 
obtained with the echelle modes of STIS, which provide the following 
resolutions: E140M $-$ $R$ = 46,000 (FWHM = 7 km s$^{-1}$); E140H with the
``Jenkins Slit'' $-$ $R$ = 200,000 (FWHM = 1.5 km s$^{-1}$).

Of course, the spectral resolution required depends on the nature of the 
measurement. For example, due to the lower mass of hydrogen, H~I absorption 
lines are often much broader than their corresponding metal lines ($b^{2} = 
2kT/m$ for thermal broadening), and consequently the resolution requirements are 
more stringent for metals than for H~I studies. Using E140M STIS data, Dav\'{e} 
\& Tripp (2001) showed that the majority of low$-z$ H~I Ly$\alpha$ lines have $b 
> 15$ km s$^{-1}$, i.e., well-resolved at E140M resolution. Therefore a 
resolving power $R = \lambda/\Delta \lambda \approx$ 20,000 is probably adequate 
for most H~I measurements. Similarly, highly-ionized species such as O~VI are 
often quite broad (e.g., Tripp et al. 2000; Tripp 2002) and may be sufficiently 
resolved at this resolution. However, the lower-ionization lines associated with 
O~VI systems are often narrow and intricate (e.g., Tripp et al. 2000; Simcoe et 
al. 2002). To study the detailed component structure of metal lines (e.g., 
column densities and line widths of {\it individual components}), resolving 
power substantially better than 20,000 is often required. It is crucial to 
examine individual components to properly constrain the conditions of the gas 
and to study abundance patterns, which sometimes show remarkable variation from 
component to component (e.g., Ganguly et al. 1998).

\begin{figure}
\plottwo{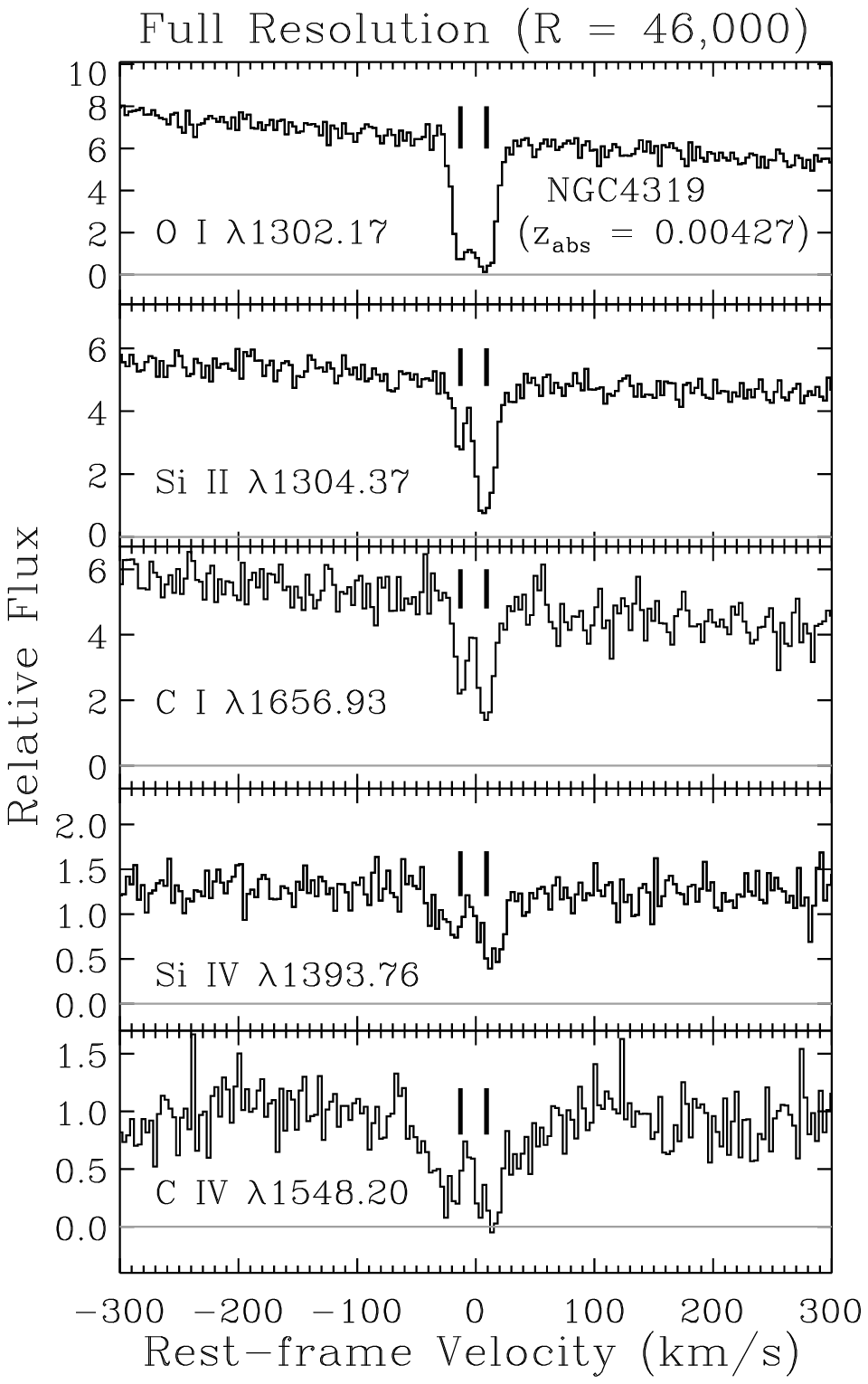}{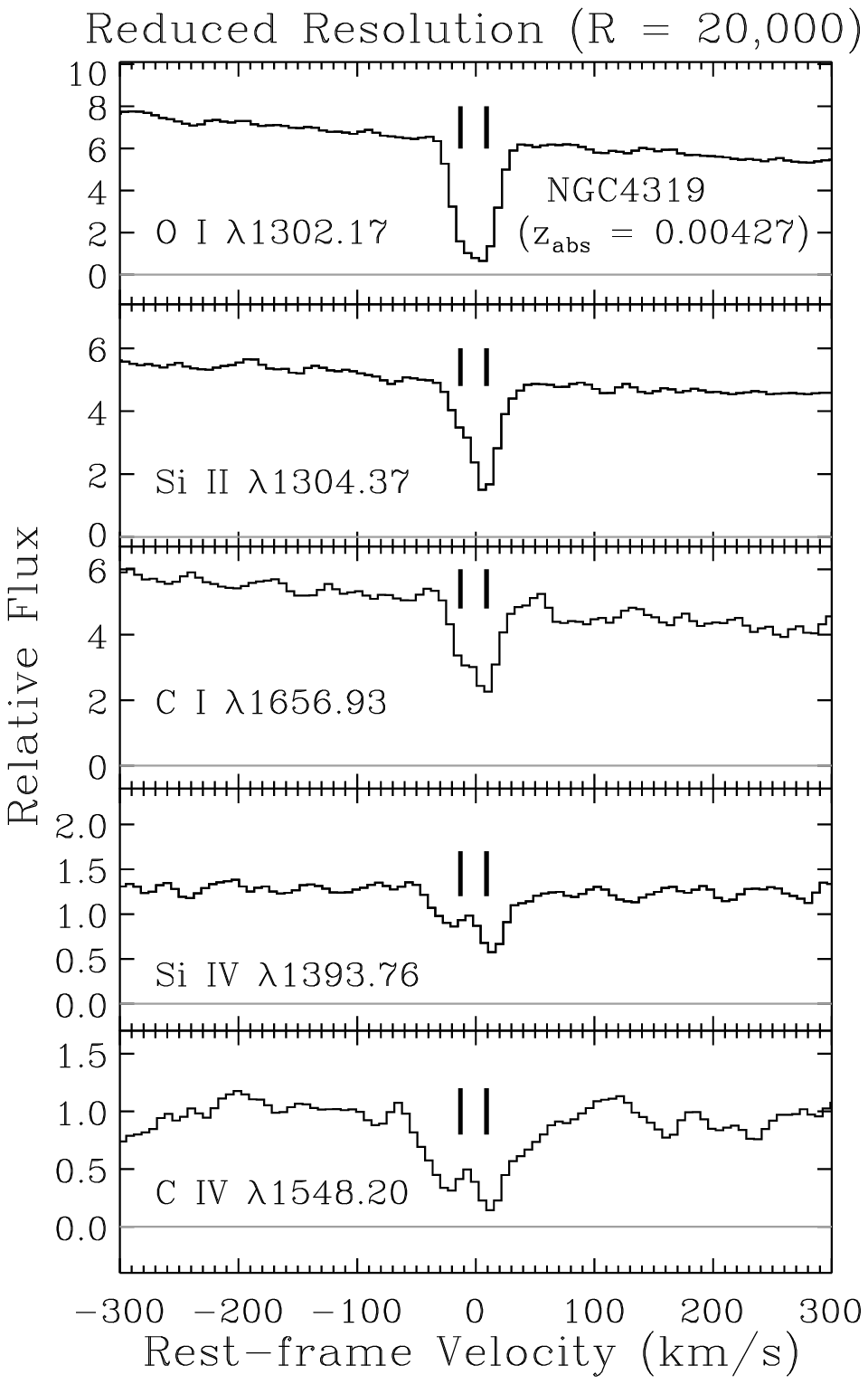}
\caption{Absorption lines of the spiral galaxy NGC4319 ($z_{\rm abs} 
= 0.00427$) in the spectrum of Mrk 205 at $R$ = 46,000 
(left panels), and $R$ = 20,000 (right panels).}
\end{figure}

\section{The ISM of NGC4319}

The sight line to the QSO Mrk 205 provides a useful example of the benefits of 
high resolution. This sightline passes through the interarm region of the 
foreground spiral galaxy NGC4319 at a projected distance of 2.8 $h_{100}^{-1}$ 
kpc from the galaxy center (see Bowen et al. 1995 and references 
therein). 
Selected absorption lines due to the ISM of this spiral, recorded at the full 7 
km s$^{-1}$ E140M resolution, are shown in the left panels of Figure 1. At 7 km 
s$^{-1}$ resolution, two components are readily apparent in the profiles of C~I, 
O~I, and Si~II. The high ion (Si~IV and C~IV) profile components are remarkably 
similar, but not identical, to those of the low ions. $R$ = 46,000 is sufficient 
to constrain the component parameters for all of the species in Figure 1 except 
O~I, where the two components are severely blended. Unfortunately, at $R$ = 
20,000, much of this information is lost, as shown in the right panels of Figure 
1. At this resolution, the O~I line appears to be a single component, and 
moreover the various species appear to have rather different component 
structure. Two components are still apparent (albeit badly blended) in Si~II and 
C~I. However, attempting to fit two components to the $R$ = 20,000 Si~II data 
leads to serious errors, as illustrated in Figure 2. The lower resolution is not 
sufficient to back out the correct parameters of the individual components. It 
is worth noting that the equivalent widths and total (i.e., integrated) column 
densities are correctly measured from the lower-resolution data, but the 
individual components are poorly constrained.

\begin{figure}
\plotone{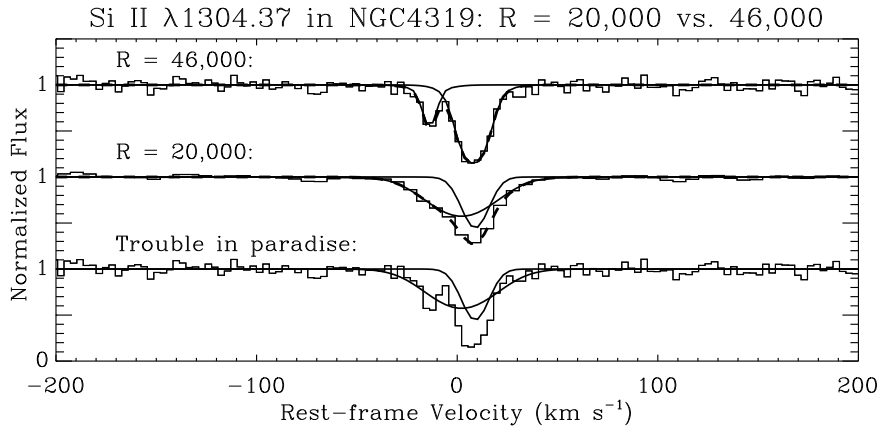}
\caption{Comparison of profile-fitting results for the Si~II $\lambda$1304.37 
transition in the ISM of NGC4319 at (top) the full $R$ = 46,000 resolving power
of the original STIS data, and (middle) the same data reduced to $R$ = 20,000. 
In each case, the data are plotted with a histogram, the two components that 
were fitted to the data are shown with solid
lines, and the combined fit is shown with a dashed line. The two components 
from the $R$ = 20,000 fit are overplotted on the full resolution data at the 
bottom of the figure; clearly the reduction in resolution led to seriously 
erroneous results.}
\end{figure}

\section{C~I in Damped Ly$\alpha$ Absorbers}

The component structure in Figures 1 and 2 is relatively simple. Figure 3 shows 
a more daunting example: Milky Way C~I lines in the sight line to HD210839 
($\lambda$ Cep). Here $R$ = 20,000 is certainly insufficient, but with $R$ = 
200,000 and coverage of many C~I lines (i.e., broad wavelength coverage), the 
C~I profiles can be effectively analyzed (Jenkins \& Tripp 2001). The $E(B-V)$ 
and $N$(H~I) are substantially higher toward $\lambda$ Cep than in NGC4319, but 
C~I lines have been detected in several damped Ly$\alpha$ QSO absorbers (e.g., 
Quast et al. 2002; Petitjean et al. 2002) that do have high H~I column 
densities. For analysis of these systems, very high spectral resolution is 
paramount.

High resolution is expensive, and it is certain that compromises between 
resolution and throughput will be required in the design of the {\it HST} 
successor, based on the science drivers.  
In our experience, the resolution provided by the E140M echelle mode of STIS
($R \approx$ 50,000) is adequate for a large fraction of the current questions
about the IGM and ISM, but STIS can only observe the brightest targets. 
Ability to go deeper at this resolution would be highly valuable.

\begin{figure}
\plotone{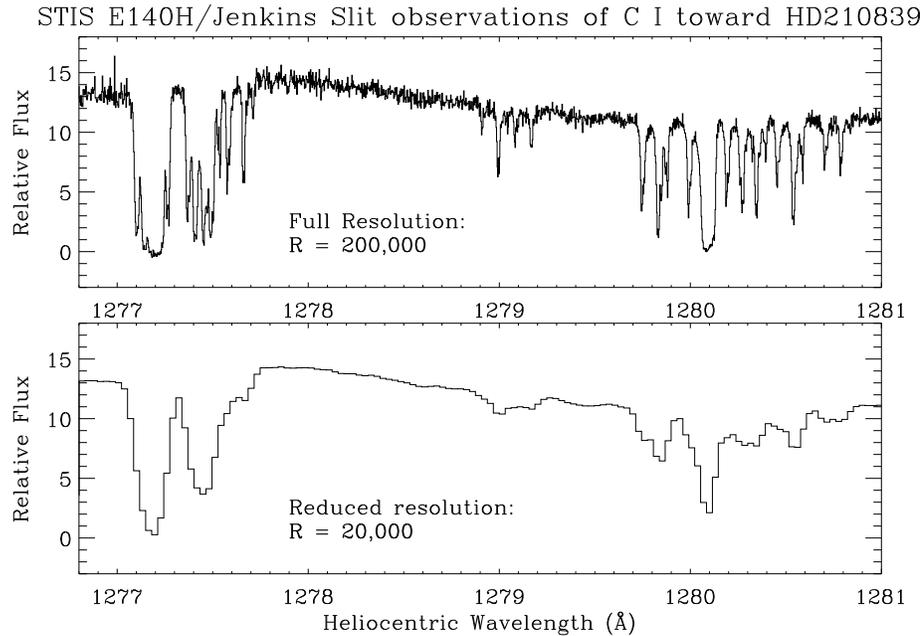}
\caption{(top) STIS E140H echelle spectra of HD210839 ($\lambda$ Cep) at 
$R$ = 200,000 (from Jenkins \& Tripp 2001). (bottom) Same data 
reduced to $R$ = 20,000.  The estimated distance to this star is 880 pc in the
direction $l = 103.8, b = 2.6; E(B - V) = 0.57$.}
\end{figure}

\end{document}